\documentclass[fleqn,twoside]{article}
\usepackage{espcrc2}

\usepackage{graphicx}
\usepackage[figuresright]{rotating}

\newcommand{\AmS}{{\protect\the\textfont2
  A\kern-.1667em\lower.5ex\hbox{M}\kern-.125emS}}

\title{Inelastic Final State Interactions in $B \to PP$ Decays}

\author{P. \.Zenczykowski \address[IFJ]{Institute of Nuclear Physics, Polish
Academy of Sciences\\
        Radzikowskiego 152, 31-342 Krak\'ow, Poland}%
        \thanks{This work was supported in part by the Polish State Committee
	for Scientific Research grant 2 P03B 046 25.}}
       
\begin{document}

\begin{abstract}
A method parametrizing all inelastic final state interactions (FSI) 
in $B\to PP$ decays is presented ($P$ - pseudoscalar meson). 
The method explicitly shows how rescattering leads to the replacement of
the short-distance amplitudes with the effective quark diagram amplitudes,
and how it affects the extraction of the unitarity triangle angle $\gamma $ 
from the data.
It is furthermore pointed out that the size of FSI effects cannot be 
determined from $B^0_d \to K^+ K^-$ decays in a satisfactory way.
The case of SU(3)-violating FSI is also discussed.
When fits to the branching ratios of all $B \to PP$ decays are performed with
all inelastic FSI included, 
the extracted value of $ \gamma $ is shifted down by some $20^o-30^o$
when compared to the no-FSI analyses, and becomes consistent with
the Standard Model value of $ 65^o \pm 7^o$ .
\vspace{1pc}
\end{abstract}

\maketitle

\section{Introduction}

One of the objectives of contemporary studies of $B$ meson decays is to check
whether their description provided by the 
Standard Model (SM) is correct. Should the values of
the SM parameters, extracted in various ways, turned out to be inconsistent,
 we might conclude that some kind of new physics is needed. 
Since new physics is expected to enter through loop diagrams,
it should appear in penguin amplitudes.

At the same time one has to keep in mind 
that diagrams of penguin topology
 can be also generated by ordinary final state interactions (FSI).
 Thus, it is important that all effects of FSI are subtracted before
 any claim as to the presence of new physics is made. However, controlling
 rescattering effects in B decays is non-trivial as FSI may be 
 highly inelastic \cite{inel}. Indeed, it follows from our knowledge of high
 energy scattering that when two pseudoscalar mesons $PP$ collide
 at $B$-meson energy, a many-body state is generally produced.
 Therefore,
 in $B \to PP$ decays one may expect contributions 
  not only from the standard $B \to P_1P_2$ quark-level transitions (corrected 
  for the elastic $B \to P_1P_2 \to P_1P_2$, and quasi-elastic
  $B \to P'_1P'_2 \to P_1P_2$ transitions), but also from the 
  inelastic $B \to M_1M_2...M_n \to P_1P_2$
  processes with many mesons in the intermediate state.
  Taking into account all of the latter FSI effects is possible 
  only if substantial
  simplifications in their description are made.

\section{Simplified description of FSI}
Since contributions from inelastic FSI
are incalculable, the only feasible way to consider them is to express 
somehow their
effects in terms of a small number of effective parameters.
The issue of how to do that
was addressed in \cite{LZ1,ZL2}, where 
several essential
simplifications, listed below, were made.

First, it was assumed that the FSI effects lead to an only small correction 
to the standard description in terms of short-distance (SD) amplitudes. 
The FSI-corrected amplitudes $\bf W$ (a set of all relevant
amplitudes $B \to hadrons $) could therefore be written as
\begin{equation}
\label{R}
{\bf W}={\bf w}+{\bf R}{\bf w},
\end{equation} 
where $\bf w$ denotes all $B \to hadrons $ SD amplitudes and $\bf R$ is the
rescattering matrix.

Second, it was assumed that FSI is $SU(3)_F$ symmetric (with $SU(3)_F$
breaking considered later). In this way, rescattering effects in all
$B^{\pm},B^0_{d,s},\bar{B}^0_{d,s}\to PP$ decays are related.

Third, an essential simplification was made in the treatment of
many-body intermediate states which may occur in $B \to P_1P_2$
transitions. 
As the original weak decay leads initially to $q\bar{q}$ or $qq\bar{q}\bar{q}$
states which evolve into many-body states only later, e.g.:
\begin{equation}
\label{manybody}
B \to q\bar{q} q\bar{q}\to M_1M_2...M_n \to P_1P_2,
\end{equation}
it is natural to include the second transition above ($q\bar{q} q\bar{q} \equiv
M'_1M'_2 \to
M_1M_2...M_n $) into the definition of FSI. Thus, as the intermediate states only
two-body states $M'_1M'_2$ may be taken 
(the $q\bar{q}$ state 
may also be considered as decaying to $M_1M_2...M_n$ through
the $M'_1M'_2$ stage).
 
Fourth, a way to sum over all intermediate $M_1M_2$ states was proposed
(hereafter we suppress the primes in $M'_1M'_2$).
Consider e.g. a tree amplitude $T_{M_1M_2}$ for the production of
a general two-body intermediate state $M_1M_2$. Without loosing
generality one can always write:
\begin{equation}
\label{TM1M2}
T_{M_1M_2}=\eta^T _{M_1M_2}T_{P_1P_2},
\end{equation}
i.e. express the amplitude $T_{M_1M_2}$ in terms of the SD tree amplitude
$T_{P_1P_2}$.
Similarly one can always write for the penguin amplitude:
\begin{equation}
\label{PM1PM2}
P_{M_1M_2}=\eta^P _{M_1M_2}P_{P_1P_2},
\end{equation}
with analogous expressions for other diagram types. 
Since FSI effects constitute a correction, only
the dominant
SD amplitudes (i.e. $T$,$P$ and the color-suppressed $C$ amplitude
in $\Delta S =0$ decays) need to be taken into account.
The essential simplification consists here in assuming that:
\begin{equation}
\label{eta}
\eta^T_{M_1M_2}=\eta^P_{M_1M_2}=...=\eta _{M_1M_2}.
\end{equation}
The above assumption helps in reducing the number of effective FSI parameters.
Further drastic reduction in their number  is achieved
via the summation over all intermediate states $M_1M_2$.
Consider for example the contribution from tree amplitudes 
$T_{M_1M_2}$ with a part of FSI in which no flavor quantum numbers are
exchanged in the $t$-channel of the rescattering amplitude (Pomeron exchange).
Denoting the $M_1M_2\to P_1P_2$ amplitude by $f_{M_1M_2}$ 
one can then write
\begin{equation}
\label{noflavortransfer}
\sum_{M_1M_2}T_{M_1M_2}f_{M_1M_2}= R_f T_{P_1P_2},
\end{equation}
where  
$R_f \equiv \sum_{M_1M_2}\eta _{M_1M_2}f_{M_1M_2}$ was introduced
as an effective parameter.
If the part of FSI with the topology of e.g. Fig.1(u) is taken into account
one similarly obtains:
\begin{equation}
\label{flavortransfer}
\sum_{M_1M_2}T_{M_1M_2}g_{M_1M'_2}= R_g P_{P_1P_2},
\end{equation}
with appropriately defined $R_g$,
i.e. an effec\-tive penguin diagram is created (Fig. 1(p)).

\begin{figure}[ptb]
\vspace{-0.91cm}
\caption{Quark line diagrams: (u) uncrossed, (c) crossed, (p) 
penguin generation.~~~~}
\phantom{xx}
{\hspace{0.3in}\includegraphics[width=2.2in]{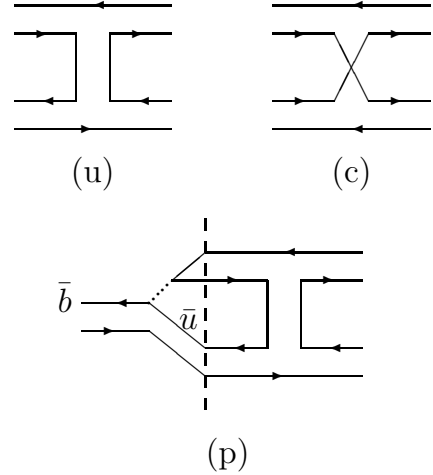}}
\vspace{-0.35cm}
\end{figure}

Fifth, Zweig rule was assumed. This means that only two types of 
(flavor transfer) FSI diagrams 
are possible as shown in Fig.1(u) (uncrossed), and Fig.1(c) (crossed).
For FSI of the uncrossed type two $SU(3)_F$ forms are possible:
\begin{eqnarray}
\label{uplus}
&Tr(\{M^+_1,M^+_2\}\{P_1,P_2\})u_+(M_1M_2),&\\
\label{uminus}
&-Tr([M^+_1,M^+_2]\{P_1,P_2\})u_-(M_1M_2),&
\end{eqnarray}
where $u_{\pm}(M_1M_2)$ are parameters describing the strength of the relevant
transitions.
Bose symmetry requires that the final $PP$ state is described by a symmetric
form $\{P_1,P_2\}$.
Charge conjugation invariance of strong interactions requires that the product of
$C$-parities of mesons \footnote{C-parities of whole $SU(3)_F$ meson multiplets
are defined to be the C-parities of their neutral members}
 $M_1M_2$ is positive for (\ref{uplus}) and negative for 
(\ref{uminus}), respectively:
\begin{eqnarray}
C_{M_1M_2}&=+C_{P_1P_2}&=+1,\\
C_{M_1M_2}&=-C_{P_1P_2}&=-1.
\end{eqnarray}
For the FSI of the crossed type only one $SU(3)_F$ form, symmetric under
$P_1 \leftrightarrow P_2$, can be written:
\begin{equation}
\label{c}
Tr(M^+_1P_1M^+_2P_2+M^+_1P_2M^+_2P_1)c(M_1M_2).
\end{equation}
As it was shown in \cite{LZ1,ZL2},
when summation over $M_1M_2$ is performed all inelastic FSI effects are 
ultimately reduced to the
appearance of {\em three} effective FSI parameters (analogues of $R_f,R_g$) only:
\begin{eqnarray}
u_R&\equiv &u_++u_- \nonumber \\
d_R&\equiv &(u_+-u_-)/2 \nonumber\\
c_R&\equiv &c,
\end{eqnarray}
($u_{+}$ involves now sums of terms including $u_+(M_1M_2)$, etc.).
If only $P'_1P'_2$ intermediate states are admitted, it follows that $u_-=0$ and
only two parameters remain.

\section{Effective quark diagrams}
When the sums over all types of $\{M_1,M_2\}$ states 
and over all types of $[M_1,M_2]$ states are performed, one obtains expressions for
the FSI-corrected $B \to PP$ amplitudes 
${\bf W} = {\bf w}+{\bf R}(u_R,d_R,c_R){\bf w}$.
For $SU(3)_F$-symmetric FSI the obtained expressions are identical in form 
to the SD expressions, but with redefined amplitudes \cite{LZ1}.
Specifically, taking selected $\Delta S =0$ amplitudes as an example, one obtains:
\begin{eqnarray}
W(B^+\to \pi ^+ \pi^0)&= &-(\tilde{T}+\tilde{C})/\sqrt{2}\nonumber \\
W(B^0_d \to \pi ^+\pi ^-)&=&-(\tilde{T}+\tilde{P}) -(\tilde{E}+\tilde{PA})\nonumber \\
W(B^0_d \to K^+K^-)&=&\tilde{E}+\tilde{PA}\nonumber \\
W(B^0_d \to K^0\bar{K}^0)&=&-\tilde{P} -\tilde{PA},
\end{eqnarray}
where
\begin{eqnarray}
\tilde{T}&=&T+2c_RC\nonumber \\
\tilde{C}&=&C+2c_RT\nonumber \\
\tilde{P}&=&P+u_R(T+3P)\nonumber \\
\tilde{A}&=&2d_RC\nonumber \\
\tilde{E}&=&2d_RT\nonumber \\
\label{PAtilde}
\tilde{PA}&=&4d_RP.
\end{eqnarray}
Analogous formulas hold for $|\Delta S| = 1$ amplitudes $T'$, $P'$, etc.
When explicit weak phases 
(with $\lambda_k^{(d)}=V_{kd}V^*_{kb}$, $V$ being the CKM matrix)
are introduced through
$T\equiv \lambda_u^{(d)}t$, and the top-dominated penguin
$P\equiv \lambda_t^{(d)}P_t$ is assumed,
one finds that the redefined penguin $\tilde{P}$ is of the form
\begin{eqnarray}
\tilde{P}&=&\lambda_t^{(d)}P_t(1+3u_R)+\lambda_u^{(d)}tu_R\nonumber\\
&=&\lambda_t^{(d)}\tilde{P}_t+\lambda_u^{(d)}\tilde{P}_u~,
\end{eqnarray}
i.e. the "top" penguin gets rescaled, and an "up" penguin appears.

Rescattering induces also the appearance of annihilation ($\tilde{A}$),
exchange ($\tilde{E}$), and penguin annihilation ($\tilde{PA}$), 
all proportional to $d_R$.
In particular, note that the amplitude for the $B^0_d \to K^+K^-$ decay 
is proportional to $d_R$. Thus the branching ratio for this decay does not yield any
information on the remaining two FSI parameters ($u_R,c_R$) (see e.g.
\cite{GR1998}).
One should think of $u_R=u_++u_-$ as originating from one superposition of contributions
from intermediate $C_{M_1M_2}=+1$ and $C_{M_1M_2}=-1$ states, with $d_R \propto
u_+-u_-$ being due to the other superposition. Thus, the two contribution may cancel in
$d_R$, while adding in $u_R$. Only
if we knew that pseudoscalar mesons alone contribute
 in the intermediate states, would the measured
size of $B^0_d \to K^+K^-$ indeed tell us about the size of the "up" penguin
$\tilde{P}_u$.

The third FSI parameter, $c_R$, redefines the tree and color-suppressed diagrams
according to the formula:
\begin{equation}
\frac{\tilde{C}}{\tilde{T}}=\frac{\frac{C}{T}+2c_R}{1+2c_R\frac{C}{T}},
\end{equation}
which shows that an originally small size of $C/T$ could be substantially affected
by FSI of the crossed type.

\section{$SU(3)_F$ breaking}
Since in the real world $SU(3)_F$ is broken,
for the purpose of fitting the data on $B \to PP $ decays 
it is appropriate to break 
$SU(3)_F$ both in the elastic, as well as in the quasi-elastic and inelastic
contributions.
Treatment of SU(3) breaking in elastic FSI is straightforward: from total
cross-section
data on
$\pi p \to \pi p$, $K p\to K p$, etc. one can extract the relevant $SU(3)_F$-breaking
couplings of Pomeron to mesons. Thus, elastic $SU(3)_F$-breaking FSI effects in $B\to
P_1P_2$ are
fully calculable (see \cite{LZ1,ZL2}).

For the inelastic (and quasi-elastic) $M_1M_2 \to P_1P_2$ transitions one expects
the annihilation (or exchange) of strange (anti)quarks to be suppressed when compared
to analogous amplitudes with all quarks non-strange. Data on hadron-hadron
collisions at $B$-mass energy indicate that the relevant suppression factor 
$\epsilon $ is much smaller than its $SU(3)_F$-suggested value of $1$.
Setting $\epsilon \ne 1$ invalidates the use of $SU(3)_F$ formulas.
In fact, for the $SU(3)_F$ breaking case
the general FSI formulas 
do not permit quark diagrams to be redefined in a way 
analogous to Eqs.(\ref{PAtilde}) and valid
simultaneously for all amplitudes. Instead, one has to use
the full form of relevant $SU(3)_F$-breaking expressions given in \cite{ZL2}. 

It was suggested that an estimate of the size of rescattering may be obtained from a
comparison of the branching ratios of $U$-spin-related decays 
$B^+ \to K^+\bar{K}^0$ 
and $B^+ \to \pi ^+ K^0$ \cite{F1998}.
In a world with $SU(3)_F$-symmetric FSI this could be done since 
in these two decays the $u$-quark
penguins generated from tree diagrams according to Fig.1(p) are of different relative
magnitudes when compared with the dominant $t$-quark penguins. Indeed, using
$P \approx \lambda P'$ and $T \approx T'/\lambda $ ($\lambda \approx 0.22$ being
the Wolfenstein parameter) one finds that
in $B^+ \to K^+ \bar{K}^0$ the ratio of the tree-generated penguin  to the original
penguin is $u_RT/P \approx T'u_R/P'\cdot 1/\lambda^2$, while
for the $B^+ \to \pi ^+ K^0$ decay the relevant ratio is simply $T'u_R/P'$.
With $1/\lambda^2 \approx 20 $, the FSI effects should be much more pronounced
in $B^+ \to K^+\bar{K}^0$.
However, when $SU(3)_F$ breaking in FSI is taken into account the FSI effects
in $B^+ \to K^+\bar{K}^0$ become suppressed by a factor of $\epsilon $ (with the
creation of a new $s\bar{s}$ pair being unlikely), and the
overall difference between the size of rescattering effects in the two modes
becomes proportional to $\epsilon /\lambda^2$, which may be much closer to $1$ than
$1/\lambda^2$.

\section{Fits to $B \to PP $ branching ratios}
In the actually performed fits to the $B \to PP$ branching ratios the following
sixteen channels were considered: 3 $\pi \pi$ channels, 4 $\pi K$,
3 $K\bar{K}$, 2 $\eta K$, 2$\eta ' K$, $\eta \pi^+$, and $\eta' \pi^+$.
The fits consisted in minimizing the $\chi ^2$ function involving 
theoretical and experimental branching
ratios $\cal{B}$ and their experimental errors $\Delta {\cal{B}}$:
\begin{equation}
\label{chi2}
\chi ^2 = \sum _1^{16}\frac{|{\cal{B}}_i(exp)-{\cal{B}}_i(the)|^2}
{|\Delta {\cal{B}}_i)exp)|^2}.
\end{equation}

Four SD parameters were used: $|T|$, $P'=-|P'|$, the singlet penguin $S'$, 
and $\gamma $. 
The remaining amplitudes were related by 
$T'=\frac{V_{us}}{V_{ud}}\frac{f_K}{f_{\pi}}T$, and 
$P=-e^{-i\beta}\left|\frac{V_{td}}{V_{ts}}\right|P'$  (with $\beta = 24^o$),
and by $C=\xi T$, $C'=T'(\xi -(1+\xi)\delta _{EW}e^{-i \gamma})$, with $\xi
=0.17$ and including the dominant electroweak penguin with $\delta_{EW}=0.65$.
Prior to the inclusion of FSI, all amplitudes were assumed to have weak phases
only.

Fits performed for the case of  SU(3)-breaking pure elastic FSI 
showed the latter to be negligible, and led to $\gamma$ around $100^o$
(as in the case with no FSI at all). 
In all fits involving free inelastic FSI parameters (in addition to $|T|, P'$,
etc.),
 the value of $d_R$ was set to $0$, and
maximal SU(3) breaking was assumed ($\epsilon =0$).
Then, two simplified cases were studied first: 1) with vanishing $c_R$
and free complex $u_R$, and 2) with vanishing $u_R$ and real $i c_R$ 
(this 
restriction takes care of direct-channel no-exotics condition 
for the crossed-type diagrams.) 
Results of the fits are shown in Fig.2. One can see that 
substantial shifts in the
fitted value of $\gamma$, away from the no-FSI value of around $100^o$,
 are obtained.
\begin{figure}[ptb]
\vspace{-0.85cm}
\caption{Fits with inelastic FSI:
1) dashed line - $u_R$ free, $c_R=0$;  
2) solid line - $u_R=0$, $c_R$ free.~~~~}
\vspace{0.6cm}
\includegraphics[width=2.7in]{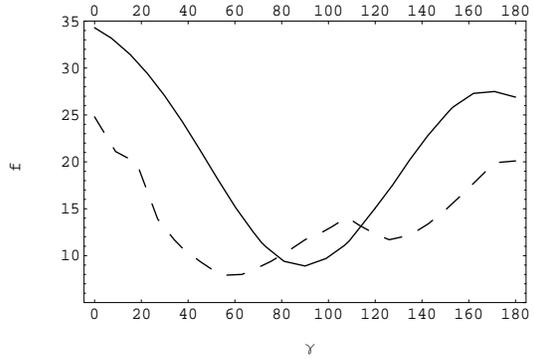}
\vspace{-0.8cm}
\end{figure}
In a general fit performed with both $u_R$ and $c_R$ free,
a very shallow global
minimum was found close to the one obtained in case 1) above.
Thus, the general fit permits values of $\gamma $ in a broad range, including
its standard model value $\gamma _{SD} \approx 65^o$ (for more details see
\cite{ZL2}). 

\section{Charming penguins}
One might question whether FSI can indeed shift the value of $\gamma $ that
much.
However, shifts of similar size are obtained also
when performing fits for charming penguins only, i.e. without
inelastic rescattering involving intermediate states composed of light quarks.
Namely, in \cite{BF1995} it was shown that one should expect the charming
penguin contribution $P_c$ to be comparable with $P_t$:
\begin{equation}
\label{BFestimate}
0.2 < \left| \frac{P_c-P_u}{P_t-P_u} \right| < 0.5.
\end{equation}
Results of the fits (shown in Fig.3) 
performed with $P_c \ne 0$ and $P_u =0$
for an updated set of branching ratios indicate \cite{Z2004}
that for the charmed
penguin of size expected in (\cite{BF1995}), 
the shift in $\gamma $ may well be of the order of $20^o$
degrees.
\begin{figure}[ptb]
\vspace{-0.85cm}
\caption{Fits for charming penguin contribution only: 
$P_c/P_t=0.6,0.4,0.0,-0.6$ for short-dashed, long-dashed, solid, 
and dotted lines
respectively}
\vspace{-0.2cm}
\includegraphics[width=2.7in]{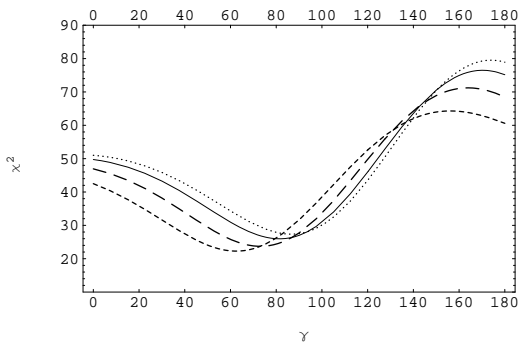}
\vspace{-1.2cm}
\end{figure}

\section{Conclusions}
In standard short-distance approaches with built-in $SU(3)_F$-symmetry the
amplitudes for B-meson weak decays may be decomposed into several quark diagram
amplitudes.
When $SU(3)_F$-symmetric final state interactions are taken into
account, the resulting FSI-corrected amplitudes must also be decompasable into
several {\em effective} quark diagram amplitudes.
In this talk the connection between the two sets of quark diagram amplitudes 
was presented for the appropriately simplified but still general case of 
inelastic FSI.
Thus, an understanding of how FSI redefine original quark diagram amplitudes was
reached.

In particular it was shown that all leading effects of 
$SU(3)_F$-symmetric FSI may be parametrized
by only three effective parameters.
Experimental upper bound on the 
 $B^0_d \to K^+ K^-$ branching ratio limits the size of one of these
 parameters only.
 The remaining two parameters redefine the penguin amplitude and mix
 the tree and color-suppressed amplitudes.
 
From the fits to sixteen branching ratios of $B \to PP$ decays
performed with no FSI taken into account it follows that
 such fits are still quite sensitive to the uncertainties present 
 in the experimental data (with the fitted value of $\gamma $ changing
 from around $100^o$ to $80^o$). 
 The fits 
performed with maximal $SU(3)_F$ breaking in inelastic FSI 
(not including charmed intermediate states) 
admit a strong shift in the extracted
value of $\gamma $ (within the range of $\gamma \in (60^o,110^o)$),
favoring the SM value of $\gamma$ very weakly.
Similar shifts and conclusions
are obtained from the fits when only charming penguins are
considered.

\end{document}